\documentclass[superscriptaddress,article,prl,showpacs,10pt,twocolumn,nofootinbib]{revtex4-1}
\usepackage{amsmath,xcolor,amsfonts,amsthm,amssymb,graphicx, srcltx,hyperref,ulem}
\usepackage{amsfonts}
\usepackage{amsmath}
\usepackage{epsfig}
\usepackage{xcolor,todonotes}
\usepackage{ulem} 
\usepackage{siunitx}
\newcommand{\beq}{\begin{equation}}
\newcommand{\eeq}{\end{equation}}
\newcommand{\ket} [1] {|#1\rangle}

\newcommand{\braket}[2]{\langle #1 | #2 \rangle}

\renewcommand{\emph}[1]{{\it #1}}

\usepackage{nicefrac}
\usepackage{graphicx,subfigure}

\usepackage[utf8]{inputenc}
\usepackage[T1]{fontenc}

\begin{document}
\title{Two-dimensional distributed-phase-reference protocol for quantum key distribution}
\author{Davide Bacco}
\email{dabac@fotonik.dtu.dk}
\author{Jesper Bjerge Christensen}
\author{Mario A. Usuga Castaneda}
\author{Yunhong Ding}
\author{Søren Forchhammer}
\author{Karsten Rottwitt}
\author{Leif Katsuo Oxenløwe}
\affiliation{Technical University of Denmark, Department of Photonics, 2800 Kgs.~Lyngby, Denmark.}

\begin{abstract}
\noindent Quantum key distribution (QKD) and quantum communication enable the secure exchange of information between remote parties. Currently, the distributed-phase-reference (DPR) protocols, which are based on weak coherent pulses, are among the most practical solutions for long-range QKD. During the last 10 years, long-distance fiber-based DPR systems have been successfully demonstrated, although fundamental obstacles such as intrinsic channel losses limit their performance. Here, we introduce the first two-dimensional DPR-QKD protocol in which information is encoded in the time and phase of weak coherent pulses. The ability of extracting two bits of information per detection event, enables a higher secret key rate in specific realistic network scenarios. Moreover, despite the use of more dimensions, the proposed protocol remains simple, practical, and fully integrable.
\end{abstract}

\maketitle

\section*{Introduction}
\vspace{-0.5cm}
\noindent Sharing sensitive information has always been a great challenge within our society. In particular, QKD, first introduced by Bennett and Brassard, provides a unique procedure for exchanging a private key, based on the laws of quantum mechanics~\cite{BBPr}. During the last decade, the effort from the scientific community has been focused on an enhancement of the quantum communication performances in terms of key rate, transmission distance and security aspects~\cite{Vallone2015c,Korzh2015,Bacco2013,Ji,Takesue2015a,Mirhosseini2015a,Usuga2010}. 
In later years this technology has matured enormously, but the lack of compact, efficient, inexpensive, and reliable systems, has restricted wide spreading of practical QKD systems.

\noindent The basic idea behind QKD systems, in the case of "prepare and measure" schemes, is based on quantum states prepared by Alice (the transmitter) and sent through a quantum channel towards Bob (the receiver). Depending on the quantum measurement, Bob can deduce which state was prepared by Alice. This way, after error reconciliation and privacy amplification methods established in a classical channel, the two users share an identical bit sequence. Ideally, QKD systems are secure with no chance for an eavesdropper to extract information on the key. However, in real implementations of the systems, due to the losses and imperfections of devices, the secret key rate defines a bound on how much information can be assumed secure~\cite{Gisin2002, Scarani2009,GBrassard2000}.

\noindent We here propose a new QKD protocol, which we refer to by the name: Differential phase time shifting (DPTS). In its essence, the protocol utilizes two degrees of freedom --- time and phase --- to encode information in a quarternary alphabet, i.e.~$\left \lbrace 0, 1, 2, 3 \right\rbrace $\cite{Usenko2005}. The DPTS belongs to the family of distributed phase-reference (DPR) protocols, which rather than using the principle of random basis-choices between different mutually unbiased bases, encodes information in adjacent weak coherent pulses ~\cite{Gisin2002,Inoue2003,Stucki2005}. We study the performance of the DPTS protocol using infinite-key analysis in the case of collective attacks, and further show that the protocol holds great potential in intracity network scenarios.

\begin{figure*}[ht]
	\center \includegraphics[width=17cm]{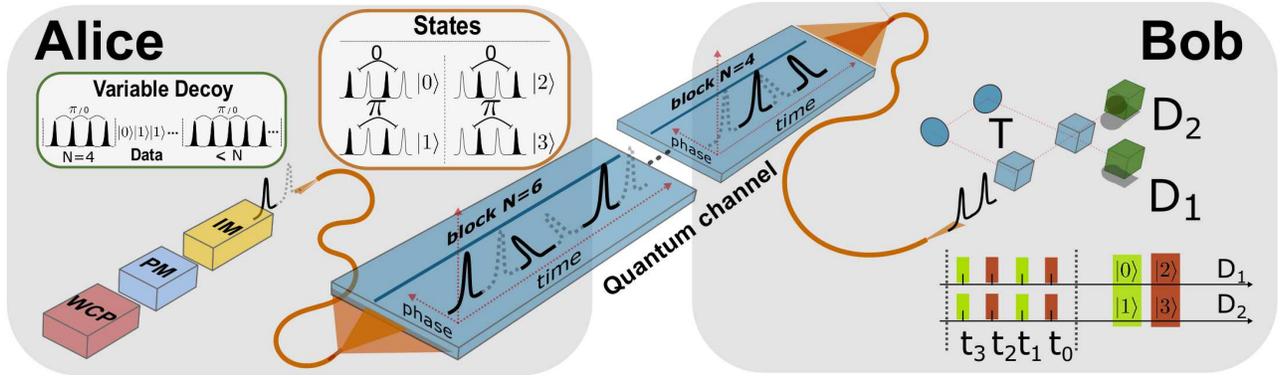}
	\caption{\textit{Basic scheme of the DPTS protocol}. A train of weak coherent pulses (WCP) is emitted by a laser of repetition rate $\nu$  ($2/T$), and attenuated to the single photon level. A phase modulator (PM) encodes the first key bit in non adjacent pulses choosing a random phase between $0$ and $\pi$. An intensity modulator (IM), exploiting two different time positions, encodes the second key bit by randomly choosing between the time instances  $ \ket{\pm \alpha} \ket{ \mathrm{vac} } $ or $ \ket{\mathrm{vac}} \ket{\pm \alpha }$. The length of the block ($N$), in which the IM uses the same time sequence, is defined by Alice who randomly decides between different duration ($N \geq 4$). In this way Alice prepares a sequence of different states: $ \ket{0}$, $ \ket{1}$, $ \ket{2}$, $ \ket{3}$. A random decoy sequence is implemented in order to check the coherence between pulses. Using a delay line interferometer ($T$ delay between arms), the receiver, Bob, can simultaneously measure the phase and the time of arrivals of the photons.}
	\label{fig:setup}
\end{figure*}

\section*{Results}

\subsection*{Principle of DPTS}
As in most practical implementations of QKD, the DPTS protocol, which is sketched in Fig.~\ref{fig:setup}, uses a source of weak coherent pulses to establish a key of random numbers between two authenticated parties, Alice and Bob. 
To initiate the key distribution process, Alice randomly encodes information in the train of pulses in two dimensions, time and phase. \textit{The time encoding} is performed using an intensity modulator (IM) as in the coherent-one way (COW) protocol \cite{Stucki2005}. For every pair of pulses (we refer to such a pair as a \textit{sub-block}), one pulse is transmitted with mean photon number $\mu < 1$ ($\ket{\alpha}$), and one is blocked completely ($\ket{\mathrm{vac}}$). Hence, within each sub-block, information is carried by the time-of-arrival of a non-empty pulse~\cite{Stucki2005}. \textit{The phase encoding} is performed using a phase modulator (PM), where a random phase between sub-blocks is either $\lbrace 0 , \pi\rbrace $. By combining the effect of the IM and the PM, Alice prepares states from the quaternary alphabet:
\begin{equation}
	\begin{split}
		\ket{0} &= \ket{\pm \alpha } \ket{\mathrm{vac}} \ket{\pm \alpha} \ket{\mathrm{vac}}, \\
		\ket{1} &= \ket{\pm \alpha } \ket{\mathrm{vac}} \ket{\mp \alpha} \ket{\mathrm{vac}}, \\
		\ket{2} &= \ket{\mathrm{vac}} \ket{\pm \alpha}  \ket{\mathrm{vac}} \ket{\pm \alpha}, \\
		\ket{3} &=  \ket{\mathrm{vac}} \ket{\pm \alpha}\ket{\mathrm{vac}}  \ket{\mp \alpha}.
	\end{split}
\end{equation}
Bob may distinguish unambiguously between these states by employing an unbalanced interferometer which interferes adjacent sub-blocks separated by  $T = 2/\nu $, where $\nu$ is the laser repetition rate. 

It is important to note that, analogous to the differential phase shift (DPS) protocol, each sub-block may participate in defining up to two states \cite{Inoue2003}. For instance, the sequence: $\ket{\alpha} \ket{\mathrm{vac} }, \ket{- \alpha} \ket{ \mathrm{vac} },\ket{- \alpha}\ket{ \mathrm{vac} }, \ket{\mathrm{vac} }\ket{ \alpha },\ket{\mathrm{vac} } \ket{ \alpha}  $ encodes the states: $\ket{1} , \ket{0} , - , \ket{2} $. Here, the '$-$' indicates a change of the temporal sequence over the sub-block separation, in which case Bob is not able to interfere the non-empty pulses in his interferometer. Therefore, to minimize the number of unused sub-blocks (or measurements), Alice may benefit from repeating the temporal encoding over long pulse sequences (i.e.~only using $\ket{0} $ and $\ket{1} $, or $\ket{2} $ and $\ket{3}$ for long intervals). However, doing so permits a potential eavesdropper, Eve, to gain partial information on a given state by measuring the time-of-arrival of pulses in adjacent sub-blocks. To take into account this possibility, Alice prepares blocks of length $N$, within which the temporal sequence of empty and non-empty pulses is the same. The value of $N$, counting both empty and non-empty pulses, is for each block chosen randomly in a uniform distribution: $N \in \left\lbrace 4, 6, ... N_{max} \right\rbrace $. This modification means that both Bob and Eve are essentially unaware of the positions of the block separations, and, whereas this is of no importance to Bob, it is fundamental for Eve.

The security of DPTS relies on the same principle as other DPR protocols: the coherence between non-empty pulses \cite{Gisin2004,Inoue2005}. Eve can not perform a measurement on any finite number of states without at some point breaking coherence between successive pulses. This is specifically true for the DPTS protocol since Eve is completely ignorant about the start and end of blocks (note that coherence is not carried across a block separation corresponding to two sub-blocks of different temporal sequences). However, since coherence is distributed across sub-block separations whereas the temporal information lies within sub-blocks, a sophisticated Eve can address each sub-block separately trying to just learn the time-of-arrival information (i.e.~is a state $\ket{0}, \ket{1}$ or is it $\ket{2}, \ket{3}$). Doing so, she only breaks coherence \textit{within} sub-blocks, and thus Bob, who only checks coherence \textit{across} sub-blocks, is not able to reveal her presence. To counter this attack, Alice introduces decoy sequences with probability $p_{decoy} \ll 1 $ \cite{Gisin2004}, in which blocks consist of $N$ non-empty pulses. Interestingly, this decoy is just a DPS sequence in which the phase encoding is carried between every second pulse (as measured by Bob). 
Consequently, if Eve probes one or more sub-blocks containing two non-empty pulses, 
she inevitably disturbs the phase relation between these pulses \cite{Scarani2009}. As a result, there are cases where Eve introduces phase errors into the communication.

\subsection*{Protocol definition}
We now describe in detail how Alice and Bob establish a common key using the DPTS protocol:
\begin{itemize}
	\item Alice prepares states for transmission in the quantum channel using her phase- and intensity modulators. We assume that Alice chooses equally and randomly between the four different states $\left \lbrace 0, 1, 2,  3 \right\rbrace $.  The temporal sequence is repeated within each block of random length ($N \geq 4$), whereas the phase difference between each sub-block is changed randomly between $\lbrace 0 , \pi \rbrace$.
	\item Once Bob has received a photon in one of the two detectors, he reveals over a public classical channel the sub-time (the number of the sub-block) instances of his recorded detection events. 
	\item Alice reports back by telling which of the events corresponded to an overlap between adjacent blocks with opposite temporal sequence (a block separation was present in that instance). Bob must discard these events.
	\item For each of the remaining detection events, Alice and Bob establish two bits of information for their key: Alice easily figures out the detection time from her sent temporal sequence, and infers from her phase encoding which detector clicked at Bob's side.  
	\item After estimating the quantum bit error rate (QBER), Alice and Bob perform standard error reconciliation and privacy amplification~\cite{Buttler2003,Martin2014,bennett95}. At the end of the process Alice and Bob share secure identical keys.
\end{itemize}

\subsection*{Secret key rate}
\noindent To further describe the proposed protocol, let us consider the maximum extractable secret key rate $R_{sk}$ \cite{Scarani2009}. For the DPTS protocol this quantity reads
\begin{equation}
	R_{sk} = f R_{B}  ~ [I_{AB}-\mathrm{min} ( I_{AE}, I_{BE} ) ]  ,
	\label{eq:SecretRate}
\end{equation}
where $R_{B}=R+ 4 p_d (1-R)$ is the total detection rate with $R= \left[ 1-\exp(- \mu t \eta_d) \right]/2 $. $\mu$ is the mean photon number of non-empty pulses, $t$ represents the quantum channel transmission coefficient, $\eta_d$ is the (common) detector efficiency, and $p_{d}$ is the dark count probability. The pre-factor $f =\left(1 - p_{decoy} \right) \left(\left< N \right> -1\right)/\left< N \right>$, where $\left< N \right>$ is the average block length, takes into account the fraction of Bob's detection events that is assigned to the key string. The unused fraction $1/\left< N \right>$ is due to detections associated with adjacent sub-blocks of different temporal sequences. In these cases, the clicks are randomly distributed between the two detectors, and so the instances are discarded.

The mutual information between Alice and Bob, is expressed in terms of the Shannon entropy as $I_{AB} = H(A) - H(A|B)$ \cite{Nielsen}. Alice has a total of four different states to choose from, and by assuming that she prepares each state with equal probability, one finds $H(A) = - \sum _{i = 1} ^4 (1/4) \log _4 (1/4) = 1$. Note that we, for convenience, measure information using a base-4 logarithm rather than the common base 2 [in units of bits one acquires $H(A) = 2$]. Furthermore, the conditional entropy $H(A|B)$ is expressed as
\begin{equation}
	H(A|B) =  S_4 \! \left( 1- e_r^{(1)} \right) + \sum_{i = 2}^4 S_4 \! \left(e_r^{(i)}  \right) ,
	\label{eq:HAB}
\end{equation}
with  $S_4 (x) \equiv - x \log_4 x $ , and where the four error probabilities are given as
\begin{equation}
	\begin{split}
		e_r ^{(1)} & = \frac{ R \frac{1 - V}{2} + 3 p_d (1- R ) }{ R_B }, \\
		e_r ^{(2)} & = \frac{ R \frac{1 - V}{2} +  p_d (1-  R ) }{ R_B }, \\
		e_r ^{(3)} & = e_r ^{(4)}   = \frac{ p_d (1-  R ) }{ R_B}, 
	\end{split}
	\label{eq:errorprobs}
\end{equation}
where $V = (p_{D_1}-p_{D_2})/(p_{D_1}+p_{D_2}) $ represents the visibility of the 
interferometer used by Bob and $p_{D_1}$ ($p_{D_2}$) represents the probability of 
detection in detector $D_1$ ($D_2$). Note that, in the definition of the error 
probabilities, the  visibility appears in only two of the four terms, since an interferometer error does not alter the time of arrival.  As a result, the DPTS protocol is less effected by interferometer mismatches as compared to the DPS protocol. On the other hand, the higher dimensionality of the DPTS protocol renders it more vulnerable to dark counts (one dark click produces two errors on the key), effectively limiting its use at long communication distances.

In order to evaluate the achievable secret key rate for Alice and Bob, we next introduce an upper bound on the information that a potential eavesdropper might obtain by performing the most basic attack; the beam-splitting attack. A complete analysis would concentrate on $I_{BE}$ since Eve is clueless about detection events resulting from imperfections at Bob's side [see Eq.~\eqref{eq:SecretRate}]. However, as a first attempt to estimate her information, we restrict ourselves to the more simple analysis of $I_{AE}$.

\subsection*{Security analysis}
This section presents an analysis of security based on the collective beam-splitting attack (BSA) and follows the method used in~\cite{Branciard2008} for the DPS and COW protocols. In the BSA, Eve replaces the quantum channel connecting Alice and Bob by a lossless line. Using a beam-splitter to simulate the losses of the quantum channel, Eve acquires $1-t$ of the signal without disturbing the state sent by Alice. Thus, the BSA belongs to the family of zero-error attacks, and is therefore undetectable by Alice and Bob. The states prepared by Alice consist of sequences $ \bigotimes _k \! \ket{ \alpha_k }$ with $ \alpha_k  \in  \left\lbrace + \alpha,0, - \alpha  \right\rbrace $, so by performing the BSA, Eve receives states of the form $ \bigotimes _k \! \ket{ \alpha_k^{(E)} }$, where $\alpha_k^{(E)}  \in   \left\lbrace +\alpha_E,0, - \alpha_E \right\rbrace $ with $\alpha_E = \alpha \sqrt{1-t}$.

\begin{figure}[ht]
	\begin{center}
		\subfigure[]{
			\includegraphics[scale=0.57]{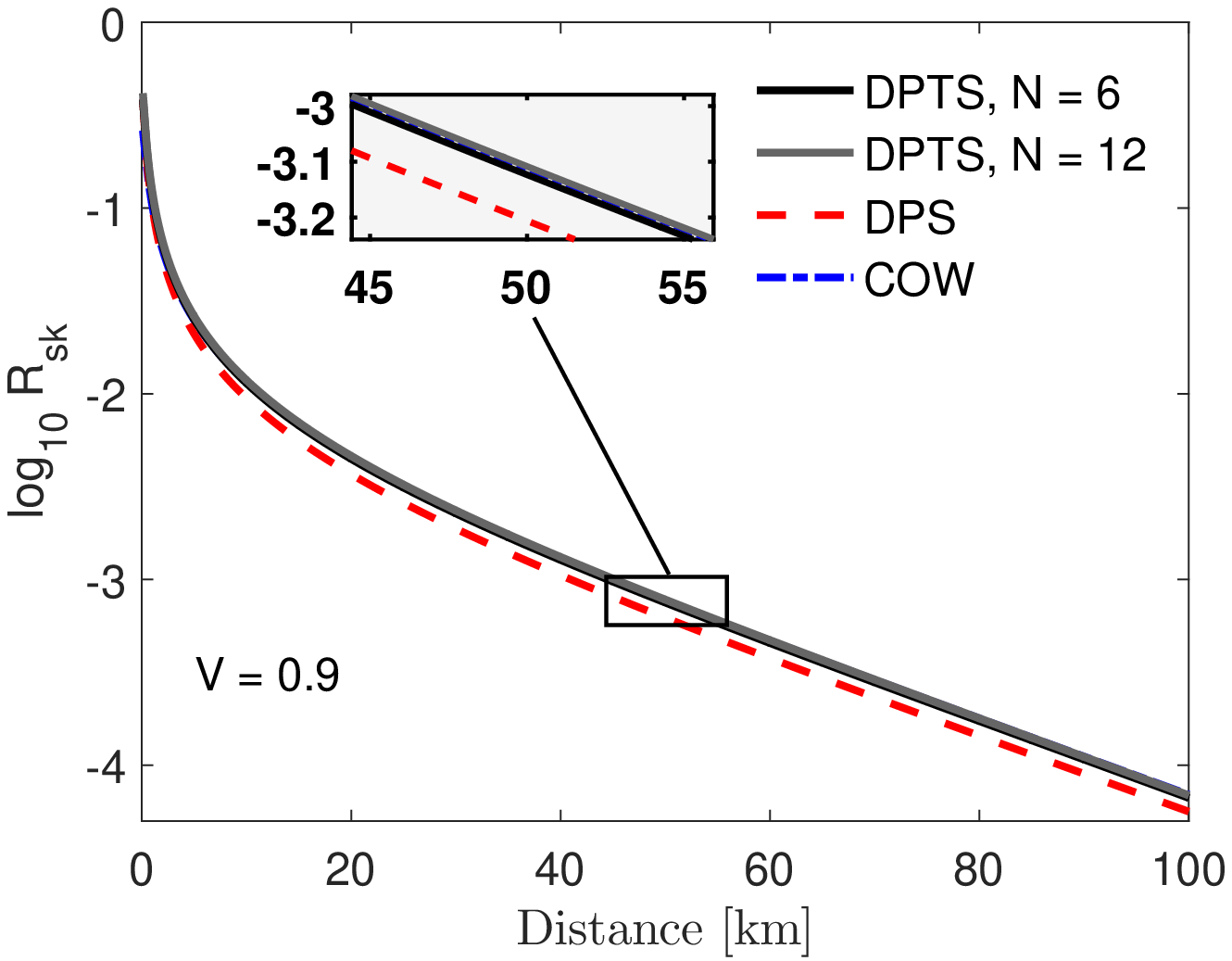}
			
		}
		\subfigure[]{
			\includegraphics[scale=0.57]{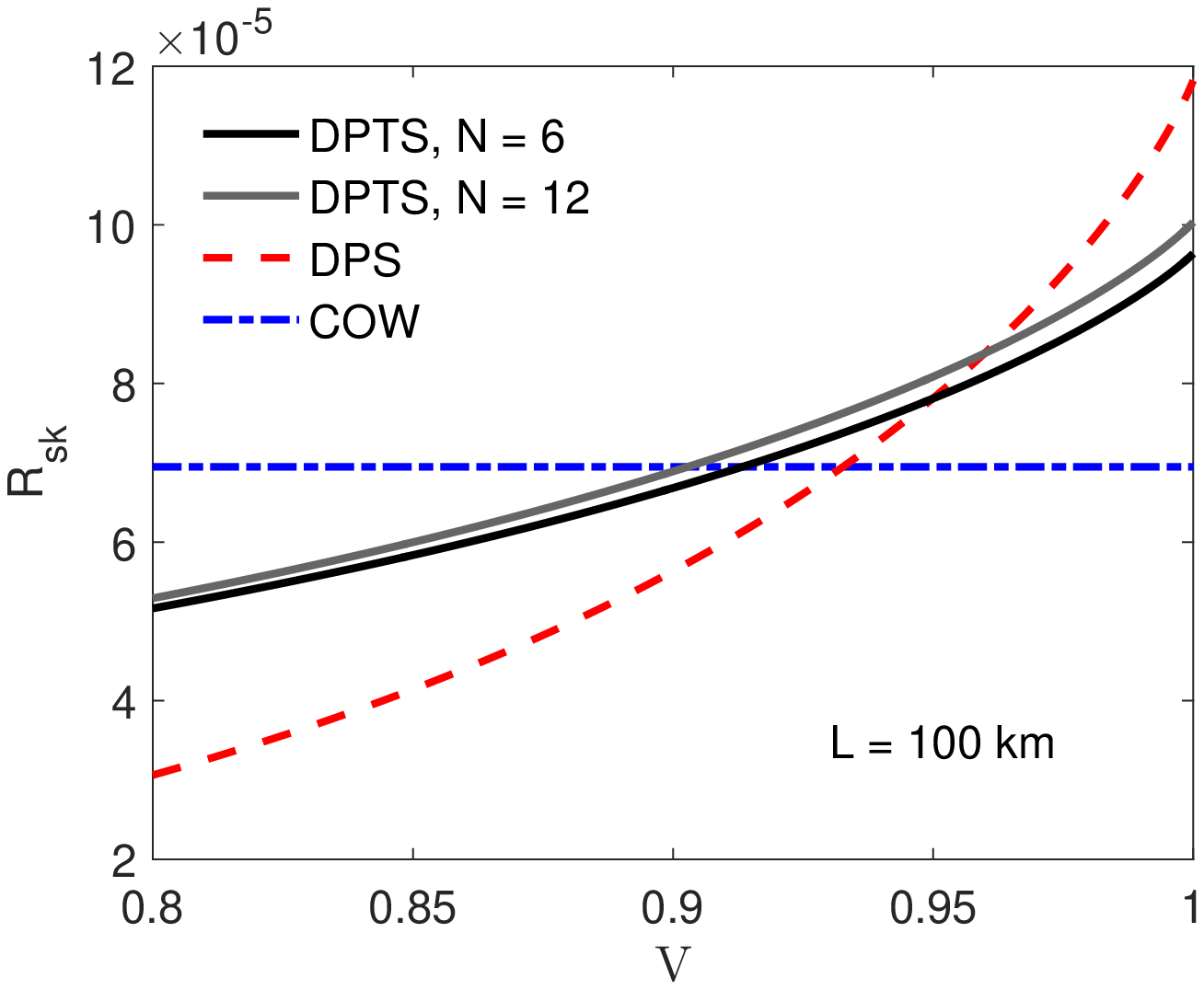}
		}
	\end{center}
	\caption{\textit{Secret key rate per pulse}. Performance versus a) distance in the case of fixed visibility, $V= 0.9$, and b) visibility at a channel length of $L = 100~\mathrm{km}$. For each of the three protocols, an optimization was performed with respect to the mean photon number $\mu$ (see supplementary material). Parameters: $\eta_d = 0.1$, $p_d=10^{-7}$, $\alpha_{loss}=0.2~\mathrm{dB/km}$, and $p_{decoy = 0.02}$ for COW and DPTS.}
	\label{fig:Rsk}
\end{figure}

At this point we assume that Eve stores the states in her quantum memory for measurement after Bob reveals his detection events. Indeed, for such a collective attack, the maximum information she may extract is given by the Holevo quantity (which must be maximized with respect to the strategies available to Eve, though here we only consider the BSA)~\cite{Scarani2009,Devetak2005}
\begin{equation}
	\chi_{AE} = S\left( \rho_E \right) - \sum _j p_j S\left( \rho_{E|j} . \right)
	\label{eq:ChiAE}
\end{equation}
Here $S$ is the von Neumann entropy, $\rho_E = \sum_j p_j \rho_{E|j}$, $p_j$ is the probability of Alice preparing the four states $j \in \left \lbrace 0, 1, 2, 3 \right\rbrace $, and $\rho_{E|j}$ is Eve's state conditioned on preparation of state $j$. As mentioned earlier, we consider only the balanced situation where Alice prepares each state with a probability $p_j = 1/4$.
In the current protocol each value in the quaternary alphabet is encoded in four consecutive pulses. It follows that Eve's states conditioned on Alice's preparation are
\begin{equation}
	\begin{split}
		\rho_{E|0} =&\frac{1}{2} \left(P_{+\alpha_E, \mathrm{vac} , + \alpha_E,\mathrm{vac}} +  P_{-\alpha_E, \mathrm{vac} , - \alpha_E,\mathrm{vac}}  \right), \\
		\rho_{E|1} =&\frac{1}{2} \left(P_{+\alpha_E, \mathrm{vac} , - \alpha_E,\mathrm{vac}} +  P_{-\alpha_E, \mathrm{vac} , +\alpha_E,\mathrm{vac}}  \right), \\
		\rho_{E|2} =&\frac{1}{2} \left(P_{\mathrm{vac},+\alpha_E, \mathrm{vac} , + \alpha_E} +  P_{\mathrm{vac},-\alpha_E, \mathrm{vac} , -\alpha_E}  \right), \\
		\rho_{E|3} =&\frac{1}{2} \left(P_{\mathrm{vac},+\alpha_E, \mathrm{vac} , - \alpha_E} +  P_{\mathrm{vac},-\alpha_E, \mathrm{vac} , +\alpha_E}  \right) ,
	\end{split}
\end{equation}
where $P_x$ is the projection operator. To calculate the maximum accessible 
information for Eve, it is 
helpful to define $\gamma =  \mathrm{e}^{-\vert \alpha_E \vert ^2}$. By this convention the overlaps between states can be written as $\vert \braket{ + \alpha_E, \mathrm{vac}, + \alpha_E, \mathrm{vac} | - \alpha_E, \mathrm{vac}, - \alpha_E, \mathrm{vac} } \vert  = \gamma^4$, and $\vert \braket{ j | k } \vert  =  \gamma^2$ for $j \neq k$, where $j, k\in \left\lbrace 0, 1, 2,3 \right\rbrace$. From this, the Holevo quantity [Eq.~\eqref{eq:ChiAE}] becomes
\begin{equation}
\begin{split}
\chi_{AE} ^{(0)} = & -\frac{(1 + \gamma ^2 )^2 + (2\gamma )^2}{8} \log _4 \left[ \frac{(1 + \gamma ^2 )^2 + (2\gamma )^2}{8} \right] \\
& - \frac{3 \left( 1 - \gamma ^2 \right) ^2 }{8} \log_4 \left[ \frac{ \left( 1 - \gamma ^2 \right) ^2 }{8} \right] \\
&-  \frac{1-\gamma ^4}{2} \log _4 \left( \frac{1-\gamma^4}{8} \right)  + h_4 \left( \frac{1 - \gamma ^4}{2} \right) .
\end{split}
	\label{eq:Holevo}
\end{equation}
where $S_4$ is defined below Eq.~\eqref{eq:HAB}, and $h_4 (x) = S_4 (x) + S_4 (1-x)$. Equation \eqref{eq:Holevo} presents an upper bound on the information Eve can obtain by trying to distinguish between the four different states. However, Eve can do better than this by trying to establish \textit{partial} information about the state Alice and Bob agreed upon.
Specifically, by performing measurements on the sub-blocks which are temporally adjacent to the time slots in which Alice and Bob agreed on the bit pairs, Eve may with some probability infer that the state was either of the pairs $\ket{0}, \ket{1}$ or $\ket{2}, \ket{3}$ (in this case, Eve has no way of knowing the phase-related bit). Since this additional attack by Eve is conditioned on her \textit{not} getting a conclusive result in the primary measurement, the corrected Holevo quantity becomes
\begin{equation}
	\chi_{AE} = \chi_{AE}^{(0)} + (1 - \chi_{AE}^{(0)} ) \: \chi_{AE} ^{(1)},
	\label{eq:HolevoCor}
\end{equation}
where $\chi_{AE}^{(1)} $ is derived and given in the supplementary material. Note however, that Eve is essentially ignorant about the position of block separations. Therefore making conclusions from this secondary attack will result in errors for Eve. 

\begin{figure}
	\begin{center}
		\includegraphics[scale=0.47]{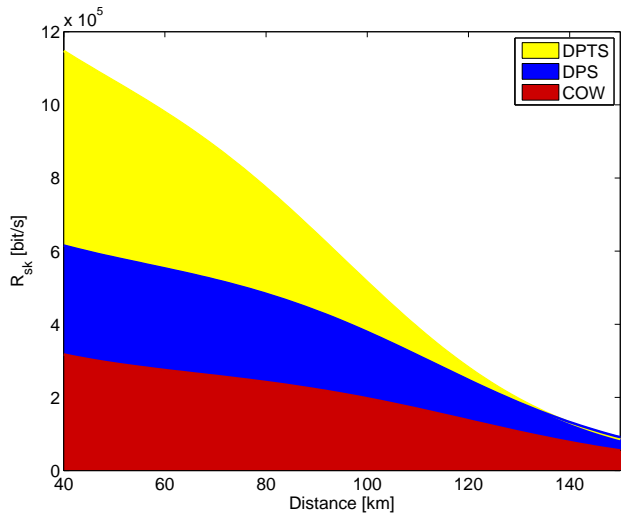}
		\caption{\textit{Secret key rate in real case scenario.} Different secret key rates achievable in a medium-length link scenario, where the detector dead times play an important role. We use mean photon numbers for the different protocols of $\mu_{DPTS}=0.23$, $\mu_{DPS}=0.19$, and $\mu_{COW}=0.52$, at repetition rate $\nu = 10 \cdot 10^9~\mathrm{Hz}$, and fixed block length of $N=4$. The detectors are specified by dark-count probability $p_d=3.5 \cdot 10^{-9}$, a dead time of $t_d=1 \cdot 10^{-6}~\mathrm{s}$, and efficiency $\eta_d = 0.1$. We assume $V=1$, and a decoy-sequence probability of $p_{decoy} = 0.02$ for COW and DPTS.}
		\label{fig:dead}
	\end{center}
\end{figure}

\subsection*{Numerical results}
\noindent Combining the results of the previous sections [in particular Eqs.~\eqref{eq:SecretRate}, \eqref{eq:errorprobs}, and \eqref{eq:HolevoCor}] enables us to plot a first upper bound on the secret key rate under the assumption of collective attacks. Specifically, Fig.~\ref{fig:Rsk} shows $R_{sk}$ versus communication distance at the optimized values of the mean photon number $\mu$. To assess the performance of the DPTS protocol, we have included plots for both COW and DPS. In comparison, the DPTS protocol has a similar performance as the other protocols under the realistic condition of non-ideal visibilities (as examples we have used $V = 0.9$ and $V = 0.95$). Noteworthy, the DPTS protocol displays a less critical dependence on the visibility when compared to the DPS protocol.

\noindent In a more realistic situation, the comparison of the protocols must take into account the detector dead times. For example, considering the case of commercial InGaAs infrared single-photon detectors (the most used in fiber links and the most promising thanks to the non-cryogenic requirement), they generally exhibit a dead time in excess of $1~\mathrm{\mu s}$~\cite{Hadfield2009,Tosi2014}. Thus, in any scenario where the detector dead time significantly influences the key generation rate, the ability to extract two bits of information per detection event grants the DPTS protocol an advantage. To illustrate this effect, Fig.~\ref{fig:dead} shows an example of the secret key rate in $\mathrm{bits~s^{-1}}$, after inclusion of the dead-time dependency.

\section*{Discussion}
\noindent The main figure of merit in a QKD system is the achievable secret key rate.  Therefore, to asses the performance of DPTS, Fig.~\ref{eq:SecretRate} displays this quantity for DPTS in comparison with the standard COW and DPS protocols. Evidently, the comparison shows very similar behavior of the three DPR protocols. Considering more specifically the case of DPTS, the final key rate is influenced by the length of the blocks $N$ prepared by Alice. Even though a higher value of $N$ allows an increased sifted key rate, it is necessary to consider a trade-off between the length of blocks and the information leakage to Eve. 
In the case of long-distance links (in excess of $100~\mathrm{km}$), the behavior of the three protocols is maintained, but as the DPTS protocol is more severely influenced by dark count events, it is generally limited to shorter distances. On the other hand, as seen by comparing the subfigures of Fig.~\ref{eq:SecretRate}, the DPTS protocol is less dependent on the interferometer visibility. This fact permits the proposed protocol to achieve a more stable secret key generation rate in comparison with the DPS protocol.
In implementing a QKD protocol, it is necessary to consider the limitations set by the optical and electronic devices \cite{Sibson2015,Diamanti2006,Takesue2006}. An important example is the single-photon detector dead time $t_d$, which sets an upper limit on the key generation rate. This parameter is important in a short- or medium-length link scenario, where the average wait time between detection events is of the same order of magnitude as $t_d$ (which is typically on the order of microseconds).   In Fig.~\ref{fig:dead}, it is shown that DPTS may achieve a significant increase in the secure key rate at distances where the detector dead time is a limiting factor. This potential arises due to the ability of the DPTS protocol to extract two bits of information per detection event.
The use of multiple degrees of freedom in transmission of information, intuitively increases the complexity of the scheme in comparison with protocols dealing with each individual degree of freedom. Despite DPTS not being an exception to this rule of thumb, the complexity overhead in comparison to DPS or COW is not crucial. On the other hand, DPTS does exhibit two significant practical advantages. Firstly, the COW protocol requires a monitoring line to check for the presence of an eavesdropper. However, such a monitoring line is unnecessary for DPTS, as an interferometer is directly used in the data line, and hence implements the necessary coherence check. Thus, the decrease in rate related to monitoring of the data line in COW, is not a limitation for DPTS. Secondly, the stability of the interferometer over time, is a considerable challenge in implementations of the DPS protocol in non-stable environments. The performance of the DPTS protocol is inherently more resilient against fluctuating interferometer visibilities, because the temporal bit remains unaffected by such inefficiencies. This entails, that DPTS might be better suited in cases where it is difficult to maintain the interferometer visibility above a certain required operation threshold.  
Finally, DPTS can potentially play an important role in QKD networks spanning from metropolitan to intercity distances~\cite{sasa11ope,froh13nat,peev09njp}. Interestingly, the required measurement apparatus is identical to the one used in DPS, and in fact, the receiver does not need to know \textit{a priori} whether the signals arise from a DPS or a DPTS encoding. 
This compatibility suggests that a versatile network encompassing the use of both the DPS and DPTS protocols is feasible. 

\noindent In conclusion, we have proposed a novel kind of distributed-phase-reference protocol for quantum key distribution. Utilizing both the time- and phase degrees of freedom, this protocol provides a significant step towards realization of fast, reliable, and practical quantum communication. Future directions include a finite-key analysis and a real-time field implementation.

\section*{Acknowledgements}
\noindent We would like to thank Dr. Giuseppe Vallone and Dr. Davide.~G.~Marangon of Department of engineering information (DEI), University of Padova for the useful discussions and for the insightful comments.
\\ \noindent Our work was supported by the DNRF Research Centre of Excellence, SPOC (Silicon Photonics for Optical Communications), ref. DNRF123.

\section*{Author contributions statement}
\noindent D.B conceived the work. D.B., J.B.C, M.A.U. and Y.D. obtained the conceptional main results. J.B.C provided security proof. J.B.C, S.R. and D.B formulated information theory analysis. K.R and L.K.Ø supervised the project. All authors discussed the results and contributed to the final manuscript.

\cleardoublepage

\section*{Additional information}
\subsection*{Eve's additional attack}
\noindent We here explore an additional (or secondary) attack option which is available to Eve when performing the beam-splitting attack (BSA). The possibility of this additional attack, arises as Alice repeats the temporal sequence (i.e.~non-empty, empty or empty, non-empty) within each block of length $N$. To clarify, assume that Bob has a detection event in a certain time slot. Eve, wanting to know which state Alice prepared for Bob, extracts the corresponding 4-pulse state from her quantum memory, and tries to determine whether it was $\ket{0}, \ket{1}, \ket{2}$ or $\ket{3}$ (See Main Text, Security analysis). Often, Eve has an inconclusive measurement and the state of the 4-pulse system is destroyed. However, in these cases, she can extract an adjacent 2-pulse state from her quantum memory, and try to learn its temporal encoding (i.e.~is it $\ket{0}, \ket{1}$ or $ \ket{2}, \ket{3}$), which is worth 1 bit of information. Unfortunately for Eve, this bit will not always be correct: In some cases she extracts a 2-pulse state belonging to an adjacent block of the opposite temporal encoding. And, essentially for the protocol, she does not know when this is the case due to the randomized block length. 

The probability of a correct bit for Eve $p_{E}$, depends on the average block size $\left< N \right>$, and can found to satisfy (assuming a negligible fraction of decoy sequences)
\begin{equation}
	p_E = \frac{\left< N \right>-2}{\left< N \right>-1} , \hspace{0.9 cm} \left< N \right> \geq 4 , 
\end{equation}
which tends towards unity for $\left< N \right>  \gg 4 $ as intuitively expected. Since the nature of the errors are identical to those of a binary symmetric channel (BSC), we can explicitly express the correction term in Eq.~\eqref{eq:HolevoCor} as
\begin{equation}
  \begin{split}
	\chi_{AE}^{(1)} = \frac{1}{2} \Big[ 1 - h_2\left(p_E \right) \Big] \bigg [S_4 \left( \frac{1+3\gamma^2}{4} \right) \\
	+3 \, S_4 \left( \frac{1-\gamma^2}{4} \right) - h_4 \left( \frac{1-\gamma^2}{2} \right) \bigg ].
  \end{split}
\end{equation}
The pre-factor of $1/2$ enters since this attack only gives half of the state information, the factor $1- h_2 \left(p_E \right) $ is the BSC capacity, and finally the three terms in the last square bracket results from analyzing how well Eve can discriminate unambiguously between the two different temporal sequences. Note that this is not identical to the expression for the coherent-one-way protocol (see \cite{Branciard2008}), since Eve's conditioned states in our case are: $\rho_{E|\mathrm{vac}} = (P_{+\alpha_E,\mathrm{vac}} +P_{-\alpha_E,\mathrm{vac}} )/2 $  and $\rho_{E|1} = (P_{\mathrm{vac},+\alpha_E} +P_{\mathrm{vac},-\alpha_E} )/2 $.

The corrected Holevo bound presented in this section only takes into account a single additional measurement performed by Eve. In principle, this measurement may be inconclusive in which case she can extract a new 2-pulse state and perform a new measurement. Thus, a more accurate analysis does exist, but is considered outside the scope of this paper as it is not expected to have a crucial impact on the bound for $\left< N \right> \leq 8 $.

\subsection*{Mean photon number parametrization}
The secret key rate $R_{sk}$ in Eq.~\eqref{eq:SecretRate} indicates that one should always try to optimize $I_{AB} - \mathrm{min} \left( I_{AE}, I_{AB} \right)$ with respect to the free variables available. For a given transmission link, an obvious parameter to optimize is the mean photon number per pulse $\mu $. In general, Bob's detection rate increases with $\mu $, but so does Eve's probability of measuring the corresponding state. Thus, for a specific setup (QKD protocol, transmission channel, interferometer, detectors, etc.), it is expected that an optimal value, $\mu_{opt}$, exists. As an example, Fig.~\ref{fig:muV95} shows the behavior of $\mu_{opt}$ versus transmission distance. These values were, for each transmission distance, obtained by numerically finding the value $\mu_{opt}$, which optimized $R_{sk}$ [which is then shown in Fig.~\ref{fig:Rsk}]. As the DPTS protocol forces a potential to distinguish both between states $\ket{+\alpha}, \ket{-\alpha}$ (as in DPS)  and $\ket{\pm \alpha}, \ket{\mathrm{vac}}$ (as in COW) it is perhaps not surprising that the optimal value $\mu_{opt}$ for the DPTS protocol lies somewhere in between the corresponding optimal values for DPS and COW.


\renewcommand{\figurename}{Additional Figure}
\setcounter{figure}{0} 
\begin{figure}
	\begin{center}
		\includegraphics[scale=0.57]{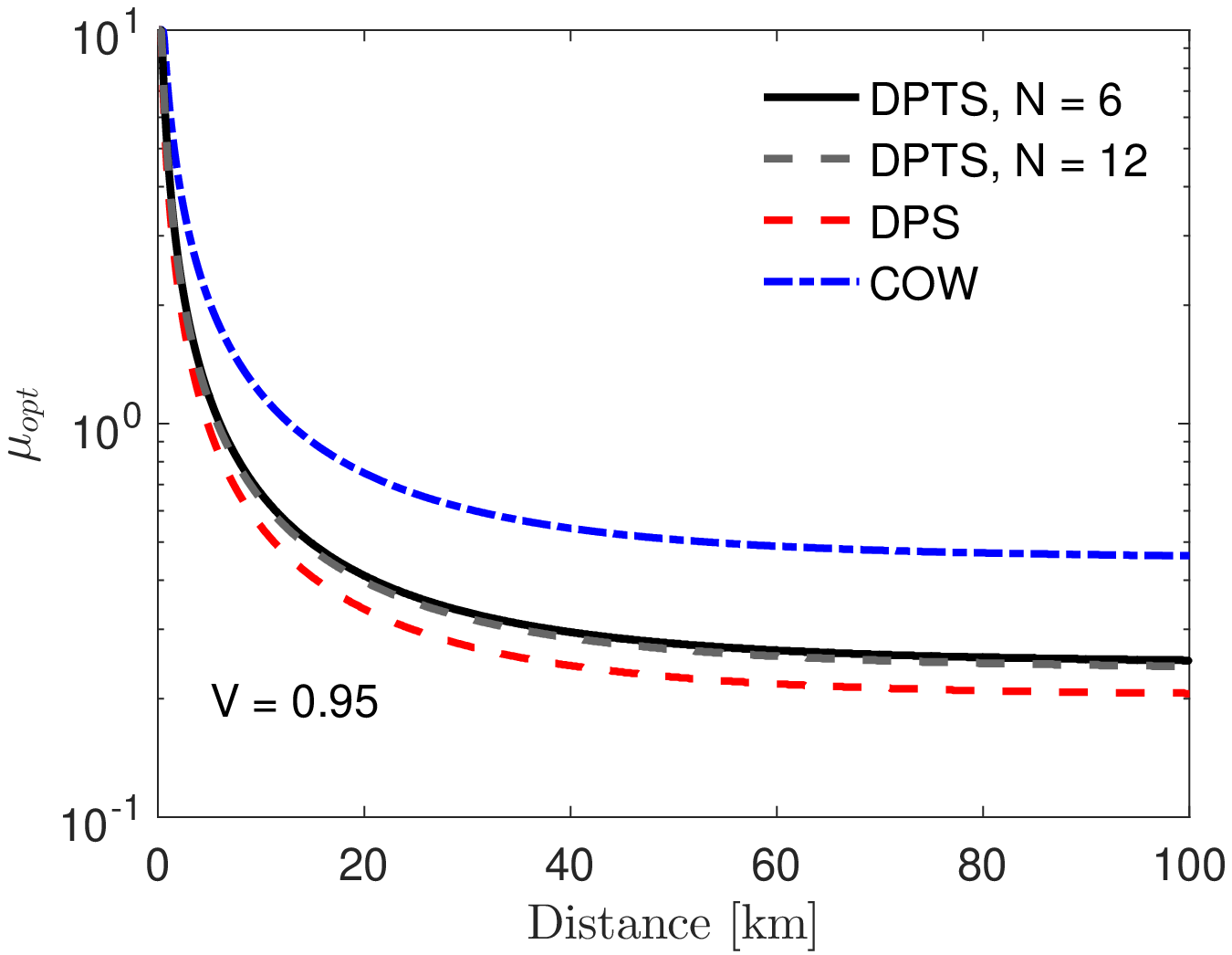}
		\caption{\textit{Optimal $\mu$ versus distance.}}
		\label{fig:muV95}
	\end{center}
\end{figure}

\end{document}